\documentstyle[12pt]{article}
\begin{document}
\begin{flushright}
{CTP-TAMU-65/94}\\
{NUB-TH-3111/94}\\
\end{flushright}
\begin{center}
{\bf THE $b\rightarrow s + \gamma$  DECAY IN SUPERGRAVITY}\\
{}~\\
R. ARNOWITT\dag\\
Center for Theoretical Physics, Department of Physics,\\
Texas A\&M University, College Station, Texas  77843-4242\\
\vskip 0.2cm
PRAN NATH\\
Department of Physics, Northeastern University\\
Boston, MA~~02115\\
\end{center}
\begin{abstract}
{\tenrm The $b\rightarrow s+ \gamma$ decay is a powerful tool for testing
models
of new physics because the new physics diagrams enter in the same loop order as
the Standard Model ones.  The current experimental and theoretical status of
this decay is reviewed.  Predictions based on the minimal supergravity model
(MSGM) in the leading order (LO) are discussed.  It is shown that results are
sensitive to the value of $m_{t}$ and $\alpha_{G}$.  The current experimental
value for the $b\rightarrow s + \gamma$ rate already very likely eliminates
part
of the SUSY parameter space when both $m_{o}$ and $m_{\tilde{g}}$ are small and
when $A_{t}$ and $\mu$ have the same sign.  Dark matter detection rates for
$\tilde{Z}_{1}$ cold dark matter for $\mu <0$ are only minimally affected by
the
current data, as are proton decay predictions for models consistent with
current
proton lifetime and $\tilde{Z}_{1}$ relic density bounds.  [\dag Invited talk
at ``Physics From Planck Scale to Electroweak Scale'', Warsaw, Sept. 21-24,
1994].}
 \end{abstract}
 \noindent
{\bf 1.~~Introduction}
\vskip 0.1cm
\indent
It is generally expected that the TeV energy domain will bring forth an array
of new physics.  There are many speculations as to what form this new physics
will take:~~supersymmetry, technicolor, compositeness, additional W and Z
bosons, etc.  Unfortunately, most of the precision LEP measurements are not
very sensitive to new physics as the Standard Model contributions enter at
the tree level, while possible new physics contributions begin at the loop
level.  Thus the most one might hope for in these measurements is a few
percent correction from new physics.

The recently discovered decay by CLEO of  $b\rightarrow s + \gamma$ is an
exception to this as this process is sensitive to new physics and is
observable for several reasons:
\begin{itemize}
\item
Being a FCNC process, it begins at the loop level so that Standard Model loops
and new physics loops enter at the same level.  Thus there can be large new
physics corrections.
\item
It is of size $G_{F}^{2}\alpha$, where $G_{F}$ is the Fermi constant (rather
then $G_{F}^{2}\alpha^{2}$ as is usual for FCNC processes)
\item
QCD corrections are large and enhance the rate by a factor $\approx$ 3.
\end{itemize}

In spite of this, it is difficult at this time to make detailed statements
concerning how well theory and experiment agree.  First the experimental
errors are still quite large.  Further, the large QCD corrections mean that
next to leading order (NLO) QCD effects are important and these are difficult
to do.  Thus theoretical errors are currently also quite large.  One can,
however, still learn interesting things at this level of knowledge, and as
experiment and theory get refined, more precise statements will become
available.\\
{}~\\
\noindent
{\bf 2.~~$b\rightarrow s + \gamma$ Decay}\\
{}~\\
\indent
The  $b\rightarrow s + \gamma$ branching ratio measured by CLEO is$^1$
\begin{equation}
BR(B\rightarrow X_{S}\gamma)=(2.32 \pm 0.5 \pm 0.29 \pm 0.32)\times 10^{-4}
\end{equation}
\noindent
where the first error is statistical, and the last two are systematic.  If one
combined errors in quadrature one obtains $BR(B\rightarrow X_{s}\gamma)\cong
(2.32 \pm 0.66)\times 10^{-4}$ which has an error of about 30\%.  In the
spectator approximation one may relate the B meson decays to the b quark
decays:\\
\begin{equation}
{{{BR(B\rightarrow X_{s}}\gamma)}\over{BR(B\rightarrow
X_{c}e\bar{\nu}_{e})}}{\cong} {{\Gamma(b\rightarrow s+\gamma)}\over {{\Gamma
(b\rightarrow c+e+\bar{\nu}_{e})}}} \equiv{R}
\end{equation}
\noindent
where $BR(B\rightarrow X_{c}e\bar{\nu}_{e})=(10.7 \pm 0.5)$\%.  [The spectator
model has corrections that begin at $0(1/m_{b}^{2})$.]

At the electroweak scale $\mu = O(M_{W})$, the elementary diagrams involve the
W-t-quark loop for the Standard Model plus additional loops (H$^{-}$-t-quark,
$\tilde{W}^{-}-\tilde{t}$-squark) for the supersymmetric generalization (Fig.
1).  Thus at $\mu \approx M_{W}$ the interaction can\\
{}~\\
{}~\\
{}~\\
{}~\\
{}~\\
{}~\\
{}~\\
{}~\\
{}~\\
{}~\\
{}~\\
\noindent
{\tenrm Fig. 1~~Elementary diagrams for $b\rightarrow s+\gamma$ decay at scale
$\mu \approx M_{W}$.  Only the third generation quark and squark contribute
significantly.}
{}~\\
{}~\\
\noindent
be described by an effective Hamiltonian$^2$
\begin{equation}
H_{eff}=V_{tb}V_{ts}^{\ast}{G_{F}\over{\sqrt{2}}} C_{7}(M_{W})Q_{7}
\end{equation}
\noindent
where $Q_{7}=(e/24\pi^{2})m_{b}\bar{s}_{L}\sigma^{\mu v}b_{R}F_{\mu v}$.
Here $F_{\mu v}$ is the electromagnetic field strength and $m_{b}$ is the b
quark mass.  However, the decay occurs at $\mu \approx m_{b}$, and one must use
the renormalization group equations (RGE) to go from $M_{W}$ to $m_{b}$.  This
produces operator mixing with the color transition magnetic moment operator
$Q_{8}=(g_{3}/16\pi^{2})m_{b}\bar{s}_{R}\sigma ^{\mu v}T^{A}b_{L}G_{\mu v}^{A}$
where $T^{A}$ and $G_{\mu v}^{A}$ are the gluon generator and field strength)
and with six 4-quark operators $Q_{1}$....$Q_{6}$.  The ratio of Eq. (2) at
$\mu = m_{b}$ becomes then
\begin{equation}
R=|{{V_{ts}^{\ast}V_{tb}}\over{V_{cb}}}|^{2}{{6 \alpha}\over{\pi I
(z)}}{|C_{7}^{eff}(m_{b})|^{2}\over{[1-{2\over 3\pi}{\alpha_{3}(m_{b})\over
\zeta}{f(z)]}}}
\end{equation}
\noindent
where $z=m_{c}/m_{b}=0.313 \pm$ 0.013,
$\eta=\alpha_{3}(M_{Z})/\alpha_{3}(m_{b})=0.548$, I(z)=1-8z$^{2}+8
z^{6}-z^{8}-24z^{4}lnz$ is a phase space factor for the $b\rightarrow
ce\bar{\nu}_{e}$ decay, and the denominator bracket (f(z)$\cong$2.41) is a
QCD correction to $b\rightarrow ce\bar{\nu}_{e}$.  In LO,
$C_{7}^{eff}(m_{b})$ is given by$^{2,3}$
\begin{equation}
C_{7}^{eff}(m_{b})={\eta^{16 \over 23}}C_{7}(M_{W})+{8\over 3}(\eta^{14\over
23}-\eta^{16\over 23})C_{8}(M_{W})+C_{2}(M_{W})
 \end{equation}
\noindent
where $C_{2}$ represents the operator mixing with the 4-quark operators.

The advantage of using the ratio R is that poorly known CKM matrix elements and
a $(m_{b})^{5}$ factor cancel out and one is left with the relatively well
known
ratios $|V_{ts}^{\ast}V_{tb}/V_{cb}|^{2}=0.95 \pm 0.04$ and $z=m_{c}/m_{b}$.
There remain, however, a number of errors and uncertainties in the above
results which we now list:\\
\begin{itemize}
\item[(i)]
Existing errors in the input parameters $\alpha_{3}(M_{Z})=0.12 \pm 0.01; z;
BR(B\rightarrow X_{c}e\bar{\nu}_{e})$; and the CKM matrix element ratio.
\item[(ii)]
Use of spectator model.  [Corrections are of O($1/m_{b}^{2}$)].
\item[(iii)]
Neglect of the next to leading order (NLO) corrections.  This is the largest
error since QCD corrections to this process are large.
\end{itemize}

An estimate of the size of the NLO corrections can be obtained by finding the
change in the LO result between running the RGE to $\mu = m_{b}/2$ and $\mu =
2m_{b}$.  Thus when higher order corrections are included, the $\mu$ dependence
should disappear, and so the $\mu$ dependence of the LO gives a measure of the
size of the NLO corrections.  The neglect of the NLO corrections is then
estimated to cause an error of about$^4~\pm~ $25\%.

Combining the above errors in quadrature, the LO calculation yield for the
Standard Model for $m_{t}$=174 GeV the result$^{4,5}$\\
\begin{equation}
BR[B\rightarrow X_{s}\gamma]\cong(2.9 \pm 0.8)\times 10^{-4}
\end{equation}
\noindent
which is about a 30\% error.  Comparing Eqs. (1) and (6), we see at this point
that it is not possible to distinguish between the Standard Model being in
agreement with experiment or differing from it by a factor of 2.  Some of the
NLO corrections have now been calculated$^{5,6}$, but a clear answer requires a
full calculation of the NLO corrections which is not easy (as they involve
calculating the finite parts of two loop and divergent parts of three loop
diagrams).

There is an additional theoretical correction due to the existence of heavy
thresholds.  (This is really part of the NLO corrections but requires special
treatment.)  The LO analysis considers the effective theory at $\mu = M_{W}$
and integrates the RGE down to $\mu \approx m_{b}$.  However, other particles
in the loop are not degenerate with W boson, and one really should start at a
higher mass scale and integrate out each particle as one crosses its mass
threshold.  For the Standard Model W-top graph, this leads to about a 15\%
enhancement$^{7,8}$.  For the SUSY case, it is estimated$^8$ that the effect is
small for the H-top loop, and perhaps $\pm$ 15\% for the $\tilde{W}-\tilde{t}$
loop.  These effects depend on the mass spectrum of the SUSY particles, and
thus
may be important as SUSY diagrams in $b\rightarrow s+\gamma$ decay can occur
with opposite sign to the Standard Model one.\\
{}~\\
\noindent
{\bf 3.  Minimal Supergravity Model (MSGM)}\\
{}~\\
\indent
In order to obtain the SUSY prediction for the $b\rightarrow s+\gamma$
branching ratio, one needs to specify the SUSY particle masses.  The MSSM is
not very useful in this respect as it depends on too many (about 20) arbitrary
parameters.  We will here make use of the minimal supergravity model$^9$ (MSGM)
which has much greater predictive power.

The MSGM has already exhibited a number of accomplishments, and has made a
number of predictions which can be tested in forthcoming experiments.  We list
some of these here:\\
\begin{itemize}
\item
The MSGM accounts for the unification of couplings at the GUT scale
$M_{G}\approx 10^{16}$ GeV implied by the LEP data$^{10}$
\item
It allows for spontaneous breaking supersymmetry in the ``hidden'' sector
(which
cannot be achieved in a phenomenologically satisfactory way in the MSSM).
\item
Flavor changing neutral interactions are naturally suppressed.
\item
The masses of the 32 new SUSY particles and all their interactions are
predicted in terms of only four parameters [$m_{o}$, the universal scalar mass;
$m_{1/2}$, the universal gaugino mass; $A_{o}$, the universal cubic soft
breaking parameter; tan$\beta \equiv <H_{2}>/<H_{1}>$] and the sign of $\mu$,
the Higgs mixing parameter.  [Here (H$_{1}, H_{2}$) are the Higgs that gives
rise to (down, up) quark masses.]  Thus the theory makes many predictions that
can be tested at$^9$ LEP2 and the LHC and at an upgraded Tevatron$^{11}$.
\item
Models with R parity yield a natural candidate for cold dark matter, the
lightest neutralino $\tilde{Z}_{1}$, with relic abundances consistent with COBE
and other astrophysical data.
\item
If representations breaking the GUT group are not large, the Gut threshold
corrections are small, and hence low energy predictions are mostly independent
of the choice of Gut group.
\item
The spontaneous breaking of supersymmetry at $M_{G}$ naturally triggers the
breaking of SU(2)$\times$U(1) at the electroweak scale by radiative breaking.
Thus the MSGM offers a natural explanation of electroweak breaking.
\item
For models with proton decay, the decay rate is suppressed by a factor of
$\approx 10^{4}$ relative to Standard Model Guts, and hence is consistent with
current data.  (However, such models generally predict rates that should be
detectable in the next round of p-decay experiments.)
\end{itemize}
{}~\\
It is convenient, in the following discussion to trade $m_{1/2}$ and $A_{o}$
for the two low energy parameters, $m_{\tilde{g}}$ (the gluino mass) and
$A_{t}$ (the t-quark A parameter at the electroweak scale).  We then explore
the entire parameter space over the range
\begin{equation}
100 GeV \leq m_{o}, m_{\tilde{q}} \leq 1~TeV;~ -6 \leq A_{t}/m_{o} \leq 6;~
tan\beta \leq 20
\end{equation}
\noindent
subject to the constraints that there be no violation of the LEP and Tevatron
bounds on SUSY masses, and that radiative breaking of SU(2)$\times$U(1)
occurs.  We chose a mesh with $\Delta m_{o} = 100$GeV, $\Delta m_{\tilde{g}}$=
25~GeV, $\Delta A_{t}$=0.5, $\Delta(tan\beta)$=2, 4.  One then determines the
masses and couplings of the SUSY particles for each of the parameter points,
which allows calculation of the $b\rightarrow s+\gamma$ rate in the LO
approximation.\\
{}~\\
\noindent
{\bf 4.~~Parameter Dependence}\\
{}~\\
\indent
The dominant contributions to the $b\rightarrow s+\gamma$ loops come from the
third generation quarks and squarks.  The $\tilde{W}-\tilde{t}$ loops play an
important role as the two stop states, $\tilde{t}_{1}$ and $\tilde{t}_{2}$
($m_{\tilde{t}_{1}} < m_{\tilde{t}_{2}}$) are very split.  The stop mass matrix
reads\\
\begin{equation}
\left({{{m_{\tilde{t}_{L}}^{2}}\atop{m_{t}(A_{t}m_{o}+\mu
ctn\beta)}}{{m_{t}(A_{t}m_{o}+\mu
ctn\beta)}\atop{m_{\tilde{t}_{R}}^{2}}}}\right)
\end{equation}
 \noindent
Because $m_{t}$ is large, the $\tilde{t}_{1}$ can become light, and in fact
large regions of the parameter space get excluded when $\tilde{t}_{1}$ is
driven tachyonic.

As is well known$^{12}$, the $H^{-}-t$ loop adds constructively to the Standard
Model to increase the $b\rightarrow s + \gamma$ branching ratio.  However, the
$\tilde{W}-\tilde{t}$ loop can enter with either sign increasing or decreasing
the total amplitude, as is seen in Fig. 2.  Note that the effect of the
$\tilde{W}-\tilde{t}_{1}$
graph is larger
for light $\tilde{t}_{1}$, as one might expect, and in the MSGM there exist
parameter points where the total branching ratio can be below or above the
Standard Model result.\\
\indent
Solving for the eiginvalues of Eq. (7) shows that $m_{\tilde{t}_{1}}^{2}$ is
smaller if $A_{t}$ and $\mu$ have the same sign or when $m_{\tilde{t}_{R}}^{2}$
is negative.  The latter can happen if $A_{t}<0$.  These effects can be seen
in Figs. (3-5)$^{13,14}$.  Fig. 3 shows that the branching ratio is largest
when\\
{}~\\
{}~\\
{}~\\
{}~\\
{}~\\
{}~\\
{}~\\
{}~\\
{}~\\
{}~\\
{}~\\
{}~\\
{}~\\
{}~\\
{}~\\
{}~\\
{}~\\
\noindent
{\tenrm Fig. 2~~Scatter plot for LO BR($b\rightarrow s + \gamma$) as a function
of $m_{\tilde{t}_{1}}$}.\\
{}~\\
\noindent
when $m_{o}$ and $m_{\tilde{g}}$ are small ($m_{\tilde{W}_{1}}\simeq ({1 \over
3}-{1 \over 4})m_{\tilde{g}})$ and
that it is larger when $A_{t}$ and $\mu$\\
{}~\\
{}~\\
{}~\\
{}~\\
{}~\\
{}~\\
{}~\\
{}~\\
{}~\\
{}~\\
{}~\\
{}~\\
{}~\\
{}~\\
{}~\\
\noindent
{\tenrm Fig. 3~~$BR(b\rightarrow s+\gamma)$ vs. $m_{\tilde{W}_{1}}$,
$m_{t}=165$
GeV, tan$\beta$=5, $|A_{t}/m_{o}|$=0.5, $\alpha_{G}^{-1}$=24.11.  Graphs (a)
and (b) are for $A_{t}<0$, (c) and (d) for $A_{t}>0$.  Graphs (a) and (c) are
for $\mu > 0$ and (b) and (d) are for $\mu < 0$.}\\
{}~\\
\noindent
 have the same sign then when they have
opposite signs (as one would expect since $m_{\tilde{t}_{1}}$ is then smaller).
The gaps in the graphs (a) and (b) occur due to the fact that the
$m_{\tilde{t}_{1}}$ has turned tachyonic (which can occur when $A_{t}$ is
negative).  Fig. 4 shows that the branching ratio increases
as $m_{t}$ increases (since $m_{\tilde{t}_{1}}$ shrinks when the off\\
{}~\\
{}~\\
{}~\\
{}~\\
{}~\\
{}~\\
{}~\\
{}~\\
{}~\\
{}~\\
{}~\\
{}~\\
{}~\\
{}~\\
{}~\\
\noindent
{\tenrm Fig. 4~~Same as Fig. 3 for $m_{t}=170$ GeV}\\
{}~\\
\noindent
diagonal elements Eq.
(7) grow) and in fact for $A_{t}<0$ the region for tachyonic
$m_{\tilde{t}_{1}}$
has grown.  Note that the $A_{t}>0$ curves are only slightly dependent on
$m_{t}$.

Fig. 5 shows the sensitivity of the branching ratio to $\alpha_{G}$.  For
$A_{t}<0$ a larger portion of the parameter space is excluded because
$\tilde{t}_{1}$ is
driven tachyonic when $\alpha_{G}$ \\
{}~\\
{}~\\
{}~\\
{}~\\
{}~\\
{}~\\
{}~\\
{}~\\
{}~\\
{}~\\
{}~\\
{}~\\
{}~\\
{}~\\
{}~\\
\noindent
{\tenrm Fig. 5~~Same as Fig. 3 with $\alpha_{G}^{-1}=24.5$}\\
{}~\\
\noindent
decreases (because $h_{t}$, the t Yukawa
coupling constant, is driven closer to its Landau pole.  (Again, for $A_{t}>0$,
there is little effect due to this small change in $\alpha_{G}$.)  If one
increases both $m_{t}$ and $\alpha_{G}^{-1}$, the effects combine to eliminate
the $A_{t}<0$ part of the parameter space for $m_{o}>100$ GeV for this case.\\
{}~\\
\noindent
{\bf 5.~~$b\rightarrow s + \gamma$ Constraints on Dark Matter Detection}\\
{}~\\
\indent
The experimental results on $b\rightarrow s + \gamma$ can impose constraints
on dark matter analyses$^{15}$.  We saw in Sec. 4, that when $\mu$ and $A_{t}$
have the same sign, the $b\rightarrow s +\gamma$ decay rate is larger than when
$\mu$ and $A_{t}$ have opposite signs$^{13}$.  This increase is particularly
marked when $m_{o}$ and $m_{\tilde{g}}$ are small.  This is also the region of
largest dark matter detection rates.  Thus one expects that for such type
situations, the experimental $b\rightarrow s + \gamma$ rate of Eq. (1) might
already be able to rule out some of the SUSY parameter space where the dark
matter detection rates are expected to be highest and hence most accessible
experimentally.

To examine this in more detail, we consider the experimental branching ratio
$BR(b\rightarrow s + \gamma)=2.32 \pm 0.66 \times 10^{-4}$ and ask, what part
of the dark matter detection rates correspond to parameter points where the
$b\rightarrow s + \gamma$ theoretical rate is within the 95\% CL of the
experimental value.  We use here the LO calculation of the $b\rightarrow s +
\gamma$ rate as a figure of merit$^{16}$.  Fig. 6 shows the maximum and minimum
event rates for a Pb dark matter detector as a function of $A_{t}$ for
$\mu<0$.  The cosmological con-\\
{}~\\
{}~\\
{}~\\
{}~\\
{}~\\
{}~\\
{}~\\
{}~\\
{}~\\
{}~\\
{}~\\
{}~\\
{}~\\
{}~\\
\noindent
{\tenrm Fig. 6~~The maximum and minimum detection rates for a Pb dark matter
detector as a function of $A_{t}$ for $\mu<0$.  The solid curve is without the
$b\rightarrow s + \gamma$ constraint while the dot-dash curve imposes the
constraint that the theoretical LO branching ratio be within the 95\% CL of the
experimental value.}\\
{}~\\
\noindent
straints on the relic
density require that the allowed parameter space be mainly for $A_{t}>0$.  Thus
for $\mu < 0$, the predicted $b\rightarrow s+\gamma$ rate is large only in the
small region where $A_{t} < 0$, and it is only in this region where the
experimental value of the $b\rightarrow s+\gamma$ branching ratio eliminates a
significant part of the parameter space.  Note that the minimum event rates are
unaffected.

Fig. 7 shows the corresponding graphs for $\mu > 0$.  Here the majority of
parameter\\
{}~\\
{}~\\
{}~\\
{}~\\
{}~\\
{}~\\
{}~\\
{}~\\
{}~\\
{}~\\
{}~\\
{}~\\
{}~\\
{}~\\
\noindent
{\tenrm Fig. 7~~Same as Fig. 6 for $\mu > 0$.}\\
{}~\\
\noindent
space occurs for $A_{t}$ and $\mu$ having the same sign, and hence the
theoretical LO $b\rightarrow s+\gamma$ decay rate is larger then the
experimental rate over most of the parameter space.  In fact, the 95\% CL bound
eliminates all of the parameter space except in the band $-0.5 \leq A_{t}/m_{o}
\leq 0.5$.\\
{}~\\
\noindent
{\bf 6.  $b\rightarrow s+\gamma$ Decay Constraints on Proton Decay}\\
{}~\\
\indent
We consider next supergravity Gut models that allow for proton decay, and
restrict the discussion to SU(5)-type models where proton decay is mediated by
a superheavy \\
{}~\\
{}~\\
{}~\\
{}~\\
{}~\\
{}~\\
{}~\\
{}~\\
{}~\\
{}~\\
{}~\\
\noindent
{\tenrm Fig. 8~~Example of SUSY proton decay diagrams where the baryon and
lepton number violations occur at the color triplet $\tilde{H}_{3}$
vertices}.\\
{}~\\
\noindent
Higgsino color triplet $\tilde{H}_{3}$ with mass
$M_{{H}_{3}}$=O($M_{G}$).  The basic diagram is shown in Fig. 8 where the
chargino $\tilde{W}$ exchange ``clothes'' the Higgsino interactions.  The decay
rate can be written in the form $\Gamma (p\rightarrow
\bar{\nu}K^{+})=const|B|^{2}/M_{{H}_{3}}^{2}$ where B represents the clothing
loop amplitude.  The current experimental bound on the proton lifetime
is$^{17}$ $\tau(p\rightarrow\bar{\nu}K^{+})>1\times 10^{32}$ yr which places a
bound on B of$^{18}$\\ \begin{equation}
|B| \stackrel{<}{\sim}100({{M_{{H}_{3}}}\over{M_{G}}}) GeV^{-1};~~ M_{G} \equiv
2\times 10^{16} GeV
\end{equation}
\noindent
In the following we limit $M_{{H}_{3}}$ to obey $M_{{H}_{3}} \leq 10 M_{G}$, so
that there not be unknown Planck scale physics entering into the analysis.

We examine here those parts of the parameter space that simultaneously
satisfies the proton decay lifetime bound and the cosmological bound that the
$\tilde{Z}_{1}$ be the cold component of dark matter, i.e.$^{19}$ 0.10
$\stackrel{<}{\sim}\Omega_{\tilde{Z}_{1}}h^{2}\stackrel{<}{\sim} 0.35$ where
$\Omega_{\tilde{Z}_{1}}=\rho_{\tilde{Z}_{1}}/\rho_{c}$ and h=H/(100 km/sec
Mpc).  Here $\rho_{\tilde{Z}_{1}}$ is the relic mass density of the
$\tilde{{Z}_{1}}$, $\rho_{c}$ is the critical mass density that closes the
Universe, and H is the Hubble constant.  We then ask what fraction of the
parameter points satisfying both the proton decay bound and the cosmological
constraints are consistent with the CLEO measurement of the $b\rightarrow
s+\gamma$ decay rate.  Using the leading order (LO) predictions for the
$b\rightarrow s+\gamma$ decay rate, we find that 95\% of the parameter points
that satisfy both the proton lifetime and cosmological bounds lie within the
95\%
CL bounds of the CLEO data.  (13\% lie within 68\% CL bounds of the CLEO data).
The parameter points satisfying the proton decay bound generally give a value
of
BR($b\rightarrow s+\gamma$) which lies above the central CLEO value.  However,
a
number of the NLO corrections$^{5}$ do decrease the theoretical value.  Thus
the
$b\rightarrow s+\gamma$ decay does not as yet seriously constrain proton decay
models.  We note that the parameter points satisfying the proton decay,
cosmological and $b\rightarrow s+\gamma$ decay bounds require\\
\begin{equation}
m_{\tilde{g}} < 375 GeV;~ m_{o} > 400 GeV;~ 0 \leq A_{t}/m_{o} \leq 0.5;~
tan\beta \leq 10
\end{equation}
\noindent
Because the proton decay bound requires tan$\beta$ to be small the dark matter
event rates will remain small, i.e.\\
\begin{equation}
R < 0.01 event/kg ~da
\end{equation}
\noindent
Thus if proton decay were seen at the next round of proton decay experiments
(e.g. at Super Kamiokande) these models would predict that $\tilde{Z}_{1}$ dark
matter would not be observable with current dark matter detectors.\\
{}~\\
{\bf 7.~~Conclusions and Summary}\\
{}~\\
\indent
The $b\rightarrow s+\gamma$ decay is a process which is sensitive to new
physics and thus is an excellent place to test models of new physics.  However,
both theory and experiment need improvement if quantitative tests are to be
made.  Using the leading order (LO) approximation to the theory, a number of
semi-quantitative results do exist:\\
{}~\\
\noindent
The rate for $b\rightarrow s+\gamma$ is largest when both $m_{o}$ and
$m_{\tilde{g}}$ are small and when $\mu$ and $A_{t}$ have the same sign.  Thus
these are the parts of the parameter space which will get eliminated if the
experimental value of the branching ratio stays below the Standard Model
number.\\
{}~\\
\indent
Large parts of the parameter space for $A_{t}<0$ gets deleted when $m_{t}$ and
$\alpha_{G}^{-1}$ increases, as the $\tilde{t}_{1}$ squark turns tachyonic.
This is a consequence of $m_{t}$ being large.\\
\indent
The current $b\rightarrow s+\gamma$ experimental bound appears to eliminate
most of the parameter space which is consistent with the $\tilde{Z}_{1}$ as
cold dark matter for $\mu >0$, but does not effect the $\mu <0$ part very
much.\\
\indent
Most of the parameter points which satisfy both current proton decay bounds and
$\tilde{Z}_{1}$ relic density bounds, are also consistent with present
$b\rightarrow
s+\gamma$ experimental rates.  However, as the data improves, the $b\rightarrow
s+\gamma$ decay will make significant impact on models that predict proton
decay.\\
{}~\\
\indent
One may expect significant constraints on new physics models once the data
and theoretical predictions on the $b\rightarrow s+\gamma$ decay become more
accurate.\\
{}~\\
{\bf Acknowledgements}\\
{}~\\
This work was supported in part by NSF grant numbers PHY-9411543 and
PHY-19306906.\\
{}~\\
{\bf References}
\begin{enumerate}
\item
E.H. Thorndike, Talk at ICHEP94, Glasgow, 1994.
\item
S. Bertolini, F. Borzumati and A. Masiero, {\it Phys. Rev. Lett.} {\bf 59}, 180
(1987); B. Grinstein, R. Springer and M.B. Wise, {\it Nucl. Phys.} {\bf B339},
269 (1990); S. Bertolini, F. Borzumati, A. Masiero and G. Ridolfi, {\it Nucl.
Phys.} {\bf B353}, 591 (1991).
\item
R. Barbieri and G. Giudice, {\it Phys. Lett.} {\bf B309}, 86 (1993); M. Misiak,
{\it Phys. Lett.} {\bf B269}, 161 (1991); {\it Nucl. Phys. Phys.} {\bf B393},
23
(1993). \item
A.J. Buras, M.Misiak, M. M\"unz, and S. Pokorski, {\it Nucl. Phys.} {\bf B424},
374 (1994).
\item
M. Ciuchini, E. Franco, G. Martinelli, L. Reina and L. Silverstrini, {\it Phys.
Lett.}~{\bf B316}, 127 (1993).
\item
M. Misiak and M. M\"unz, TUM-T31-79/94.
\item
P. Cho and B. Grinstein, {\it Nucl. Phys.} {\bf B365}, 279 (1991); Err. (in
print). \item
H. Anlauf, SLAC-PUB-6525 (1994).
\item
A.H. Chamseddine, R. Arnowitt and P. Nath, {\it Phys. Rev. Lett.} {\bf 49}, 970
(1982).  For reviews see P. Nath, R. Arnowitt and A.H. Chamseddine,
{\it ``Applied N=1 Supergravity''} (World Scientific, Singapore 1984); H.P.
Nilles, Phys. Rep. {\bf 110}, 1 (1984); R. Arnowitt and P. Nath, {\it Proc. of
VII J.A. Swieca Summer School} (World Scientific, Singapore, 1994).
\item
P. Langacker, Proc., PASCOS-90, Eds. P. Nath and S. Reucroft (World Scientific,
Singapore 1990); J. Ellis, S. Kelley and D.V. Nanopoulos, {\it Phys. Lett.}
{\bf
B249}, 441 (1990); U. Amaldi, W. de Boer and H. Furstenau,{\it Phys. Lett.}
{\bf
B260}, 447 (1991); F. Anselmo, L. Cifarelli, A. Peterman and A. Zichichi,
{\it Nuov. Cim.} {\bf 104A}, 1817 (1991); {\bf 115A}, 581 (1992).
\item
T. Kamon, J.L. Lopez, P. McIntyre and J. White, {\it Phys. Rev.} {\bf D50},
5676
(1994).
\item
J.L. Hewett, {\it Phys. Rev. Lett.} {\bf 70}, 1045 (1993); V. Barger, M.S.
Berger and R.J.N. Phillips, {\it Phys. Rev. Lett.} {\bf 70}, 1368 (1993).
\item
J. Wu, R. Arnowitt and P. Nath, CERN-TH.7316/94-CTP-TAMU-03/94-NUB-TH-3092/94.
\item
Some authors$^{4,5}$ define the LO without the denominator bracket in Eq. (4).
To be in accord with this definition one should reduce the curves of Figs.
(3-5)
by about 15\%.
\item
P. Nath and R. Arnowitt, {\it Phys. Lett.} {\bf B336}, 395 (1994);
CERN-TH.7363/94-NUB-TH-3099/94-CTP-TAMU-38/94.
\item
As discussed in Sec. 2, the LO theoretical error is about the same as the
experimental error.
 \item
Particle Data Group, {\it Phys. Rev.} {\bf D50}, 1173 (1994).
\item
P. Nath, R. Arnowitt and A.H. Chamseddine, {\it Phys. Rev.} {\bf D32}, 2348
(1985); R. Arnowitt and P. Nath, {\it Phys. Rev.} {\bf D49}, 1479 (1994).
\item
R. Arnowitt and P. Nath, CERN-TH.7362/94-CTP-TAMU-37/94-NUB-TH-3098/94.
\end{enumerate}
\end{document}